\documentclass[twocolumn,pre,showpacs,showkeys]{revtex4-1}
\usepackage{graphicx}
\usepackage{soul}
\usepackage{color}
\usepackage{amsmath}
\usepackage{pgf}
\usepackage[greek,english]{babel}
\usepackage{subcaption}
\usepackage{caption}
\usepackage{hyperref} 
\newcommand{\RNum}[1]{\uppercase\expandafter{\romannumeral #1\relax}}\usepackage{amssymb}

\begin{document}

\title{Inductively coupled Josephson junctions: a platform for rich neuromorphic dynamics} 

\author{G. Baxevanis}
\affiliation{School of Electrical and Computer Engineering, Aristotle University of Thessaloniki, 54124 Thessaloniki, Greece
}

\author{J. Hizanidis} \email[Correspondence email address:]{hizanidis@physics.uoc.gr}
\affiliation{Institute of Electronic Structure and Laser,
Foundation for Research and Technology-Hellas,
70013 Heraklion, Greece}
\affiliation{Institute of Nanoscience and Nanotechnology, National Center for Scientific Research, ``Demokritos'', 15341 Athens, Greece}

\date{\today}

\begin{abstract} 
Josephson junctions (JJs) are by nature neuromorphic hardware devices capable of mimicking excitability and spiking dynamics. When coupled together or combined with other superconducting elements, they can emulate
additional behaviors found in biological neurons. From a technological point of view,
JJ-based neuromorphic systems are particularly appealing 
since they present THz-speed processing and they operate with near-zero power dissipation.
In this work we study a system of inductively coupled JJs and focus on the nonlinear dynamical aspects of 
its neurocomputational properties. In particular, we report on spiking behavior related to a saddle-node \emph{off} invariant cycle bifurcation and excitability type \RNum{2}, synchronization, first spike latency effects, and multistability.
Special emphasis is placed on the bursting dynamics the system is capable of reproducing, and a new underlying mechanism
is proposed beyond the approach followed in prior works.
\end{abstract}



\maketitle

\section{Introduction}
Since their discovery in the early 1960s~\cite{josephson1962possible,van1981principles,LIK1986}, Josephson junctions
(JJs) remain at the forefront of advancing technology in superconducting electronics, sensing, high frequency devices, and quantum science.
An important JJ-based device is the Superconducting Quantum Interference Device (SQUID), a highly sensitive magnetometer that uses
JJs to measure extremely small magnetic fields~\cite{orlando1991foundations,hizanidis2018flux}. Josephson junctions are also incorporated in Rapid Single-Flux Quantum (RSFQ) technology,
as elements of ultrafast and low-power digital circuits~\cite{likharev1991rsfq}, and in superconducting metamaterials with unique 
tunable electromagnetic properties~\cite{cai2024effects}. 
In addition, Josephson junctions are employed in quantum computing since they constitute the key component of superconducting quantum bits (qubits)~\cite{osbourne2003superconducting}.
Another fascinating application involves the exploration of JJs
for the design of superconducting neuromorphic computing systems~\cite{schneider2022supermind}.

Neuromorphic computing with Josephson junctions presents many advantages. 
First of all, JJs are by nature neuromorphic
hardware devices capable of mimicking excitability and spiking behavior~\cite{izhikevich2000neural},
as demonstrated by the fundamental dynamical model for the device, namely the Resistively and Capacitively Shunted Junction (RCSJ) model~\cite{stewart1968current}. When combined in circuits, Josephson junctions are capable of emulating additional properties
of biological neurons. Moreover, superconductor-based 
neuromorphic systems are particularly appealing 
since they present THz-speed processing which 
far surpasses CMOS-based neuromorphic chips~\cite{milo2020memristive}. 
Furthermore, they operate with near-zero power dissipation even when cryogenic cooling is taken into account and
have excellent scaling properties.
Finally, superconducting neuromorphic systems combine classical analog dynamics with quantum effects for versatile computation.

The number of neuromorphic device implementations using superconducting elements shows a significant increase over the last few years. These involve circuit simulations, theoretical modeling, and experimental
efforts. Several works employ coupled Josephson junctions for the emulation of single neurons, transmission lines, and
synapses~\cite{crotty2010josephson,segall2017synchronization,segall2014phase,schneider2020fan,schneider2020synaptic,schegolev2023bio}
Other setups use circuitry components based on superconducting
quantum-phase slip junctions~\cite{cheng2018spiking,cheng2021high}, SQUIDs combined with JJs~\cite{mizugaki1994implementation,schneider2020fan,Karimov2024,feldhoff2024short}, superconducting nanowires~\cite{toomey2019design,lombo2021superconducting,skryabina2022superconducting}, 
or incorporate superconducting electronics with integrated photonics creating hybrid hardware platforms~\cite{shainline2017superconducting,primavera2021considerations}.
A recent review can be found in \cite{schneider2022supermind}.

As already stressed, the single Josephson junction is capable of mimicking the behavior of a biological neuron insofar 
as it exhibits excitability and spiking~\cite{izhikevich2000neural}.
When coupled together, JJs can reproduce 
even more characteristic neurophysiological properties, namely action potentials, refractory periods, and firing thresholds~\cite{crotty2010josephson}. Nonlinearity is a key feature of JJs and biological neurons
which are inherently nonlinear dynamical systems due to their ability to process, integrate,
and transmit signals in complex, time-dependent ways~\cite{dayan2005theoretical,izhikevich2007dynamical}.
Information processing in the brain depends not only on the electrophysiological properties of neurons but also on their dynamical properties.
This fact has inspired nonlinear dynamics based computing,
which exploits the rich behavior of nonlinear
dynamical systems for computation purposes~\cite{hoppensteadt1999oscillatory,Kia2017NonlinearDA}.

In this spirit, there have been a few studies on superconducting neurons
where the focus is placed on the nonlinear dynamical aspect of neurocomputation.
More specifically, Ref.~\cite{dana2006spiking} numerically confirms the spiking and bursting behavior in the Resistive Capacitive inductive Shunted 
Josephson junction (RCLSJ) model and the underlying mechanism is explained through a qualitative bifurcation analysis. 
Bursting dynamics has also been reported for resistively coupled Josephson junctions~\cite{hongray2015bursting} as well as globally coupled mixed populations of oscillatory and excitable JJs~\cite{hens2015bursting}.
The distinct form of the action potential is remarkably reproduced by a system
of inductively coupled JJs~\cite{crotty2010josephson}, a detailed bifurcation analysis confirms neural excitability type I and II, while chaos and noise-induced bursting are also observed~\cite{chalkiadakis2022dynamical}. 

In the present work, we address another configuration of inductively coupled JJs
which was first introduced in the context of superconducting interferometers~\cite{blackburn1978dynamics}, but no association was made 
with the neuromorphic properties of this device. Through an in-depth dynamical analysis
we will explore the system's rich neuron-like behavior and we will also 
attempt to elucidate the mechanisms behind certain dynamics which have been overlooked
in prior studies.

The paper is organized as follows: in Sec.~\ref{sec:model} we derive the coupled Josephson junction model 
(which from now on will be referred to as JJ neuron)
and make a brief comparison to previous relevant models. In Sec.~\ref{sec:FP} the bifurcations involving fixed points and their geometric representation are discussed. Section~\ref{sec:Dynamics} 
deals with the global dynamics and the associated neurocomputational properties of the JJ neuron, 
namely spiking behavior and excitability (Sec.~\ref{sec:spiking}), first spike latency (Sec.~\ref{sec:FSL}), synchronization (Sec.~\ref{sec:synch}), and bursting (Sec.~\ref{sec:bursting}). We summarize our results and
propose topics for further studies in Sec.~\ref{sec:Conclusions}.

\section{The Model}
\label{sec:model}
\begin{figure}[!htp]
    \centering
    \includegraphics[width=0.9\linewidth]{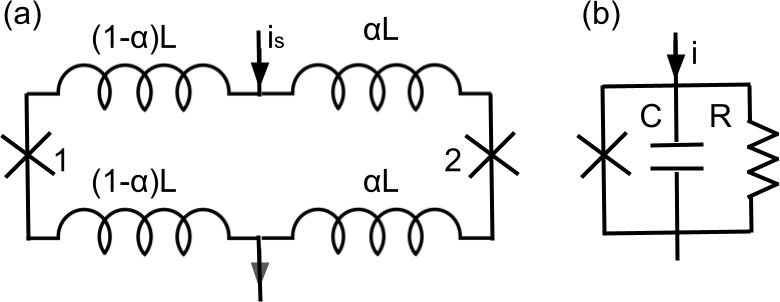}
    \caption{(a) Schematic diagram of the circuit for the inductively coupled JJ model. (b) RCSJ equivalent circuit of the single Josephson junction.}
    \label{fig:circuit}
\end{figure}

The circuit corresponding to the model under study involves two Josephson junctions in a loop, driven by a dc current source $i_s$, as shown in the schematic figure of Fig.~\ref{fig:circuit}(a), where JJs are marked with the symbol ``$\times$". The JJs are inductively coupled with the total inductance of the circuit divided into
two portions, namely $2\alpha L$ on the right and $2(1-\alpha)L$ on the left, where $\alpha$ is an asymmetry parameter in the sense that $\alpha\neq0.5$ ensures that the currents carried
by the two junctions differ. The labels ``1" and ``2" mark the left and right junction, respectively. 

The Josephson junctions are nonlinear superconducting elements made of two superconductors that are weakly coupled through a non-superconducting gap material, such as an insulator usually a
thin dielectric/oxide layer.
Each superconductor may be characterized by a single macroscopic wavefunction with a corresponding phase.
When the current applied to the JJ is less than a critical threshold $I_c$ no voltage will develop across the junction; 
that is, the junction acts as if it had zero resistance. However, the JJ is characterized by a constant phase difference
$\phi$ that satisfies the Josephson current-phase relation according to which the current is equal to $I_c\sin\phi$.
When the threshold is exceeded, a voltage develops across the junction that obeys the Josephson voltage-phase relation
$V = (\hbar/2e)d\phi/dt=(\Phi_0/2\pi) d\phi/dt$
where $\Phi_0$ is the flux quantum, $t$ denotes the time, $e$ is the electron charge and $\hbar$ is the Planck's constant.
    
Within the framework of the Resistively and Capacitively Shunted Junction (RCSJ) model \cite{LIK1986}, the JJ is
treated as a parallel circuit consisting of an ideal Josephson element, a resistor, and a capacitor, driven by a constant current source $i$, schematically
shown in Fig.~\ref{fig:circuit}(b). The current flowing through the junction 
is the sum of the supercurrent through an ideal JJ, the displacement current through the capacitor $C$ and a resistive current  through the resistor $R$:
\begin{equation}
    \frac{\hbar C}{2e}\frac{d^2\phi}{dt^2} + \frac{\hbar}{2eR}\frac{d\phi}{dt}+I_\text{c}\sin{\phi}=i.
    \label{JJ_equation}
\end{equation}
The mechanical analog of the RCSJ model is the damped pendulum driven by a constant torque.
When $i$ exceeds a critical value, a magnetic flux pulse~\cite{likharev1991rsfq} is created
in analogy to the whirling solution in the pendulum~\cite{strogatz2018nonlinear}.
This magnetic flux pulse forms the basis for the neuron-like oscillatory dynamics 
exhibited by JJ neuron model, which will be studied in the following sections. 

Using the following normalizations: $\tau^2=t^2/(LC)$, $\gamma=LI_c/\Phi_0$, $\beta=R^{-1}\sqrt{L/C}$,
$I_s=i_s/I_c$, and applying Kirchhoff's laws for the circuit of the system under study [Fig.~\ref{fig:circuit}(a)], we obtain the dimensionless equations for the JJ neuron model:
\begin{align}
    \label{eq:model_eq}
\ddot{\phi_1} + \beta \dot{\phi_1} + 2\pi\gamma \sin \phi_1 &=- \frac{1}{2}(\phi_1 - \phi_2) + 2\pi \alpha \gamma I_s, \\
\label{f2 differential}
\ddot{\phi_2} + \beta \dot{\phi_2} + 2\pi\gamma \sin \phi_2 &= \frac{1}{2}(\phi_1 - \phi_2) + 2\pi(1-\alpha) \gamma I_s,
\end{align}
where the dot notation refers to differentiation with respect to $\tau$. It is worth mentioning that the term $2\pi\gamma$ is usually referred to as the  SQUID parameter.  

In this work the focus will be on the analogies of this model to the biological neuron, i. e. we will
cover all the neuronal properties that this model is capable of emulating. The main variable whose dynamics we will
study is the voltage $V_{1,2}=\dot \phi_{1,2}$ that corresponds to the membrane potential of the biological neuron, and the
control parameters will be $I_\text{s}$ and $\alpha$ that define the postsynaptic current received by the neuron.
Note that in previous works on inductively-coupled JJs in a different circuit~\cite{crotty2010josephson,chalkiadakis2022dynamical} the action potential was reproduced by
a different quantity,
namely the sum of the two phases $\phi_1+\phi_2$, while the voltages across the two junctions corresponded to the ionic currents flowing in real neurons, $Na^+$ and $K^+$, respectively. That is, each model, depending on the coupling,
is capable of emulating neuron-like behaviors through different variables.

As far as dimensionless parameters are concerned, the values we have used stem from physically meaningful ones provided by the rapid single-flux quantum (RFSQ) circuitry~\cite{likharev1991rsfq,segall2017synchronization}. Typical values for the critical current of a single JJ are $I_\text{c} \in  [10-100\mu A ]$, for the inductance
$L\in [1-100 pH]$, the input current $i_s$ takes values close to $I_\text{c}$, and the junction size, which determines its capacitor $C$ and resistance $R$ is in the range of $0.7-5 \mu m$. Based on these values, we obtain the dimensionless parameters $\beta=4.5$, $\gamma=10$, and $\alpha=0.6$ that were also also in~\cite{blackburn1978dynamics}. 
Similarly, the bias current is kept within a plausible range $(0,2]$.

It is important to note here that the choice of $\gamma=10$ ensures large inductances, 
thus yielding the system a slow-fast dynamical model, which is crucial for the bursting behavior
that we will address in Sec.~\ref{sec:bursting}.
In the following section, we derive expressions for the fixed points 
of the system and perform a linear stability analysis. 

\section{Fixed points and geometrical representation}
\label{sec:FP}
Although fixed points in terms of neuronal dynamics
correspond to resting states and as such are not particularly interesting from a neurocomputation point of view,
we will see that they play an important role for the more complicated global dynamics presented further in the manuscript. 
By setting $V_1=\dot\phi_1$ and $V_2=\dot\phi_2$, the system of Eqs.~\ref{eq:model_eq} is tranformed to the 
following set of equations:

 \begin{align}
 \label{V1}
     \dot{\phi_1}&=V_1,\\
  \label{V1 dot}   
     \dot{V_1}&=-\beta V_1 -2\pi\gamma \sin\phi_1 -\frac{1}{2}(\phi_1-\phi_2) +2\pi \alpha\gamma I_s, \\
 \label{V2}
   \dot{\phi_2}&=V_2,  \\
\label{V2 dot}
   \dot{V_2}&=-\beta V_2 -2\pi\gamma \sin\phi_2 +\frac{1}{2}(\phi_1-\phi_2) +2\pi(1- \alpha)\gamma I_s.
\end{align}

In this way, the evolution of the system can be visualized as a trajectory in the phase plane $(\phi_1,V_1,\phi_2,V_2)$. Then, the equilibria $(\phi_1^*,0,\phi_2^*,0)$ of the system are provided by solving the following equations:
\begin{align}
\label{eq:nullcline V_1}
\phi_1^*& = \phi_2^* + 4\pi\gamma \sin{\phi^*_2} -4 \pi(1-\alpha)\gamma I_s, \\
\label{eq:nullcline V_2}
\phi_2^* &= \phi_1^* + 4\pi\gamma \sin\phi_1^* -4 \pi\alpha\gamma I_s.
\end{align}

\begin{figure}[!htp]
    \centering
      \begin{subfigure}{\linewidth}
         \includegraphics[width=\linewidth]{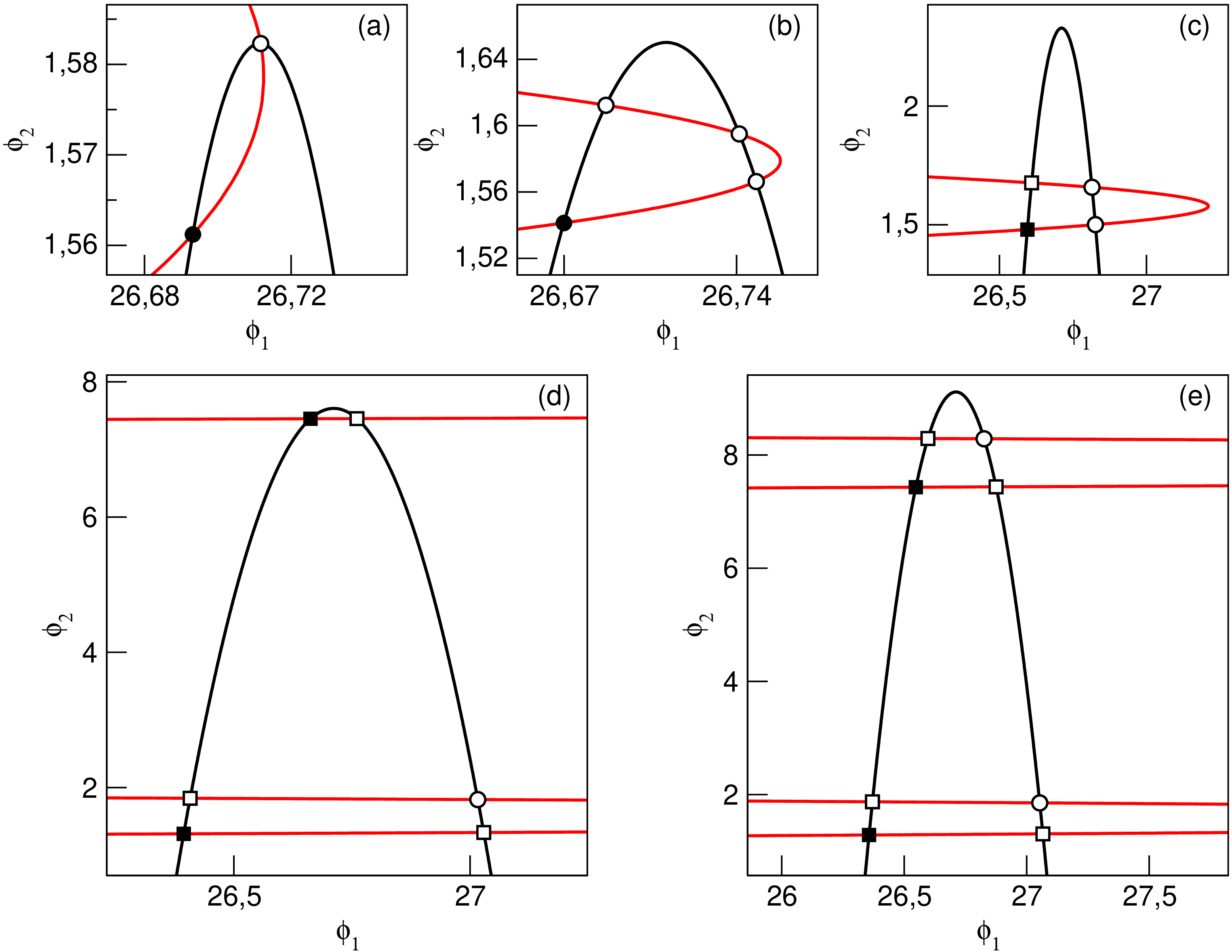}
        \phantomsubcaption{}
        \label{fig:}    
    \end{subfigure}
    \begin{subfigure}{\linewidth}
         \includegraphics[width=\linewidth]{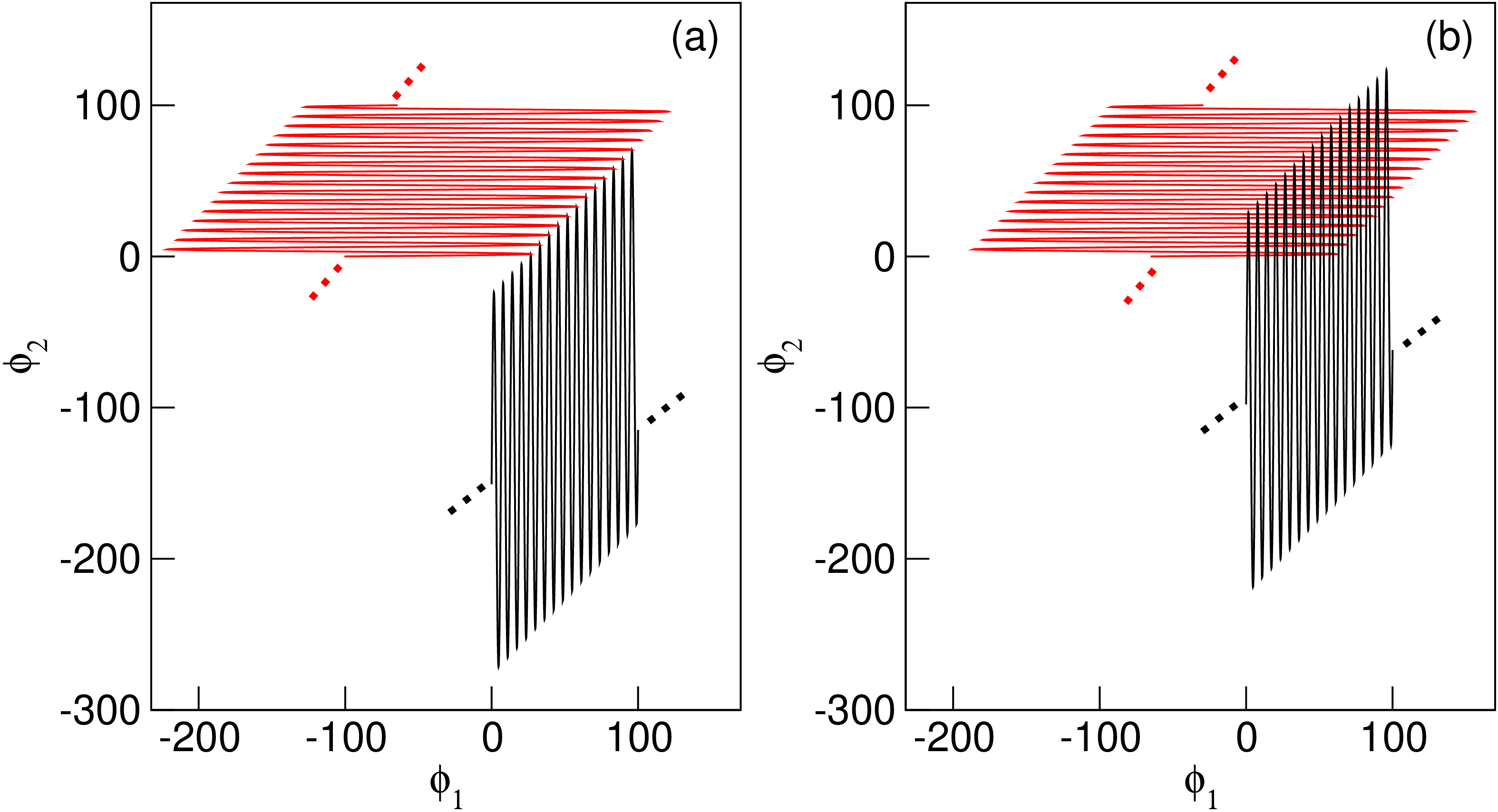}
        \phantomsubcaption{}
        \label{fig:}    
    \end{subfigure}
    \caption{Zoom-in of $V_1$ (black) and $V_2$ (red) nullclines for (a) $I_s=1.9999$, (b) $I_s=1.999$, (c) $I_s=1.99$, (d) $I_s=1.92$, and (e) $I_s=1.9$.
    Full circles denote stable nodes, open circles are saddle points, full squares are stable foci, and open squares are saddle-foci.
    Zoom-out of nullclines for (f) $I_s=2.0$ and (g) $I_s=1.3$. }
    \label{fig:nullclines}
\end{figure}
The fixed points can be found geometrically in the $(\phi_1,\phi_2)$ plane, as intersections of the $V_1$ and $V_2$ nullclines
given by Eqs.~\ref{eq:nullcline V_1}--\ref{eq:nullcline V_2}. Plugging \ref{eq:nullcline V_2} into~\ref{eq:nullcline V_1} we obtain the 
expression $\sin\phi_1+\sin\phi_2=I_s$, which automatically imposes a maximum value $I_s=2$ above which no fixed points are to be found.
This is visible in Fig.~\ref{fig:nullclines}(f) where the two nullclines are plotted in the $(\phi_1,\phi_2)$ projection,
and for this specific value of $I_s$ are tangent to each other. As $I_s$ decreases (and for all other parameters fixed) the $V_1$ nucllcline moves upward, while the $V_2$ nucllcline moves to the right, resulting in an ever increasing number of intersections,
and thus, fixed points. 

A typical case is shown for $I_s=1.3$ in Fig.~\ref{fig:nullclines}(g).
Note that the nullclines consist of infinitely many parabolic-shaped ``slices" that are equivalent to each other
since they merely are shifted by $\pm2\pi$ in both directions. Therefore, one can just focus on one of those ``slices", for instance, the ones closer to the zero horizontal axis. The maximum number of intersections (and thus fixed points) is achieved for $I_s=0$, where the overlap
of the nullclines is maximum. As $I_s$ increases, the number of fixed points obeys the empirical formula $N_{\textbf{FP}}=160-80I_s$, 
when $I_s$ is varied by a step $\Delta I_s=0.025$. For intermediate values of $I_s$ the formula does not apply per se, however, the general
rule is that the number of fixed points is always an even number which decreases
by $2$ as $I_s=2$ is approached, where finally $N_{\textbf{FP}}=0$. This happens due to the saddle-node and saddle-saddle bifurcations
that take place in turns, and will be discussed next.

\begin{figure}[h!]
     \centering
     \begin{subfigure}{\linewidth}
         \includegraphics[width=\linewidth]{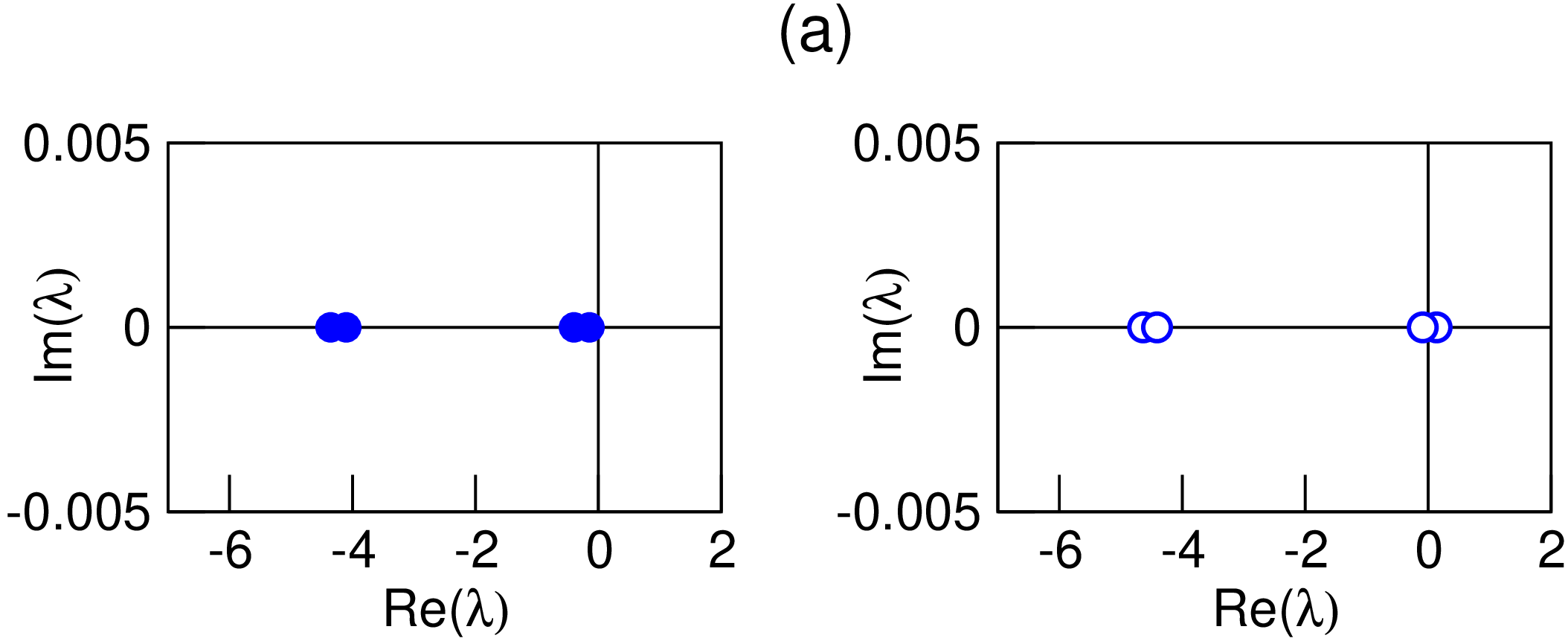}
         \phantomsubcaption{}
        \label{fig:}    
    \end{subfigure}
     \begin{subfigure}{\linewidth}
         \includegraphics[width=\linewidth]{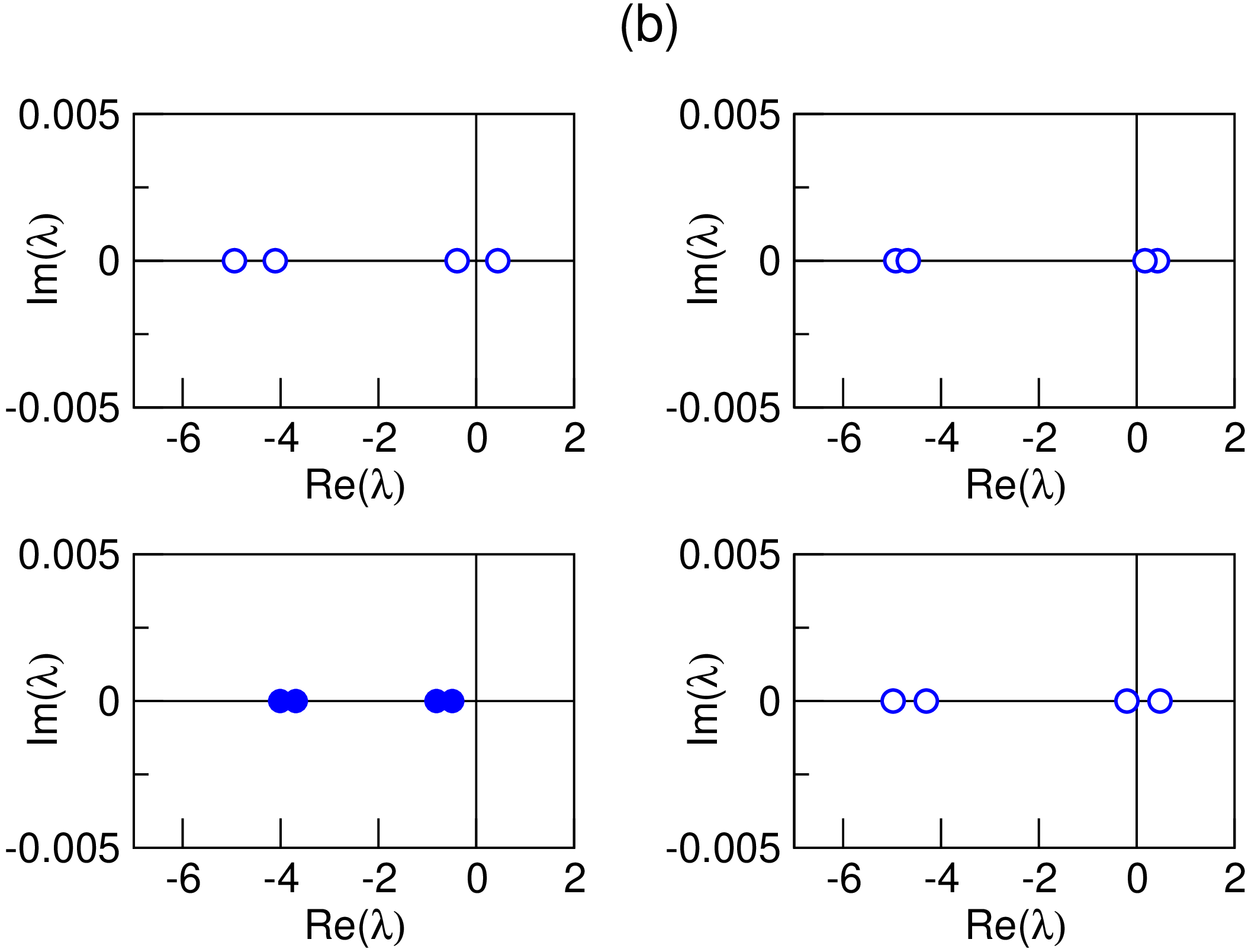}
         \phantomsubcaption{}
         \label{fig:}    
     \end{subfigure}
    \begin{subfigure}{\linewidth}
         \includegraphics[width=\linewidth]{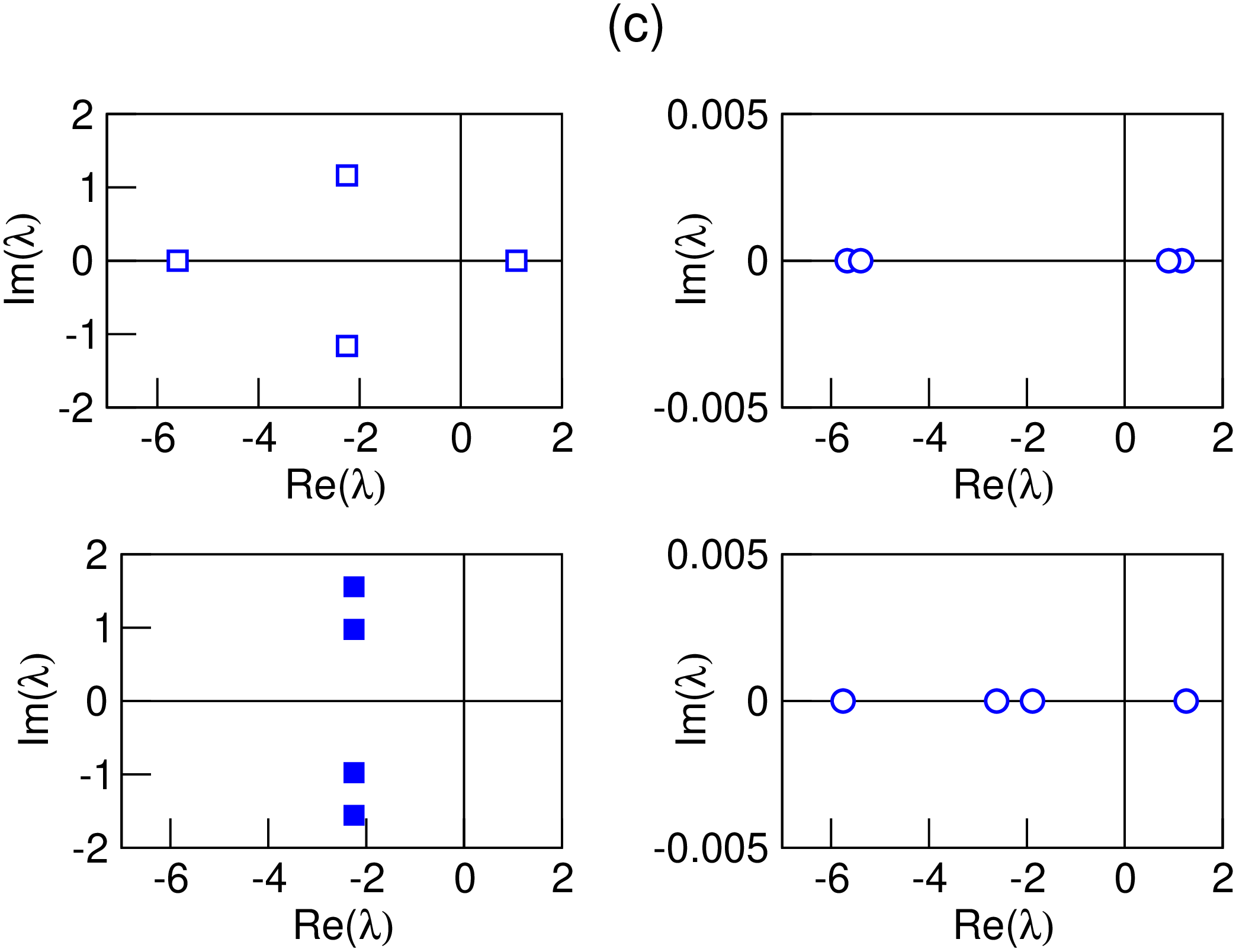}
         \phantomsubcaption{}
         \label{fig:}    
     \end{subfigure}
    \caption{Eigenvalues of the fixed points for $I_s$ values marked by labels A--C in Fig.~\ref{fig:SN_lines}(a). (a) Corresponds to point A where we have one stable node and one saddle point, (b) corresponds to B, where we have 4 fixed points (1 stable node and 3 saddle points), and (c) corresponds to point C, where we have 1 stable focus, 1 saddle-focus, and 2 saddle points.  Other parameter values: $\alpha=0.6$, $\beta=4.5$ and $\gamma=10$.}
    \label{fig:eigenvalues}
 \end{figure}
 
Our analysis will start in the vicinity of $I_s=2$ and progress downward.
Figure~\ref{fig:nullclines} shows the generation of fixed points in the interval $I_s\in[1.9,2)$. At $I_s=1.9999$ (Fig.~\ref{fig:nullclines}(a))
we have two fixed points: One stable node denoted by a full circle and one saddle point denoted by an open circle.
The stability of the equilibira can be determined via linear stability analysis (provided in the Appendix), from which 
we can calculate their corresponding eigenvalues, as shown in Fig.~\ref{fig:eigenvalues}: 
The stable node has four real negative eigenvalues shown in Fig.~\ref{fig:eigenvalues}(a) (left), while the saddle-point has three real negative eigenvalues and one positive, shown in Fig.~\ref{fig:eigenvalues}(a) (right).
These fixed points are born through a saddle-node (SN)
bifurcation. 

At $I_s=1.999$ the number of fixed points increases by 2, as explained above, namely by 2 saddle points (Fig.~\ref{fig:nullclines}(b)),
this time via a saddle-saddle (SS) bifurcation. 
At $I_s=1.99$ the number of fixed points is the same, however their quality has changed: The stable node has become a stable focus (marked by a full square), while one saddle point has turned into a saddle-focus (marked by an open rectangle), see Fig.~\ref{fig:nullclines}(c). 
The eigenvalues of the four fixed points of Figs.~\ref{fig:nullclines}(b) and (c) are shown in Figs.~\ref{fig:eigenvalues} (b) and (c), respectively. For uniformity, the symbols of the eigenvalues follow those of their corresponding fixed points. 

At $I_s=1.92$ [Fig.~\ref{fig:nullclines}(d)] the number of fixed points again increases by 2, namely by 
a pair of a stable-focus and saddle-focus, due to a SN bifurcation. Finally, at $I_s=1.9$ we have yet again an increase by 2 of the number of fixed points, this time by a pair of saddle-point and saddle-focus (Fig.~\ref{fig:nullclines}(e)), through a SS bifurcation. 
This scenario is repeated with SN and SS bifurcations occurring in turns, increasing thus, the number of fixed points until $I_s=0$ is reached, where the maximum number of equilibria is achieved.

\begin{figure}[!htp]
    \centering
   \includegraphics[width=\linewidth]{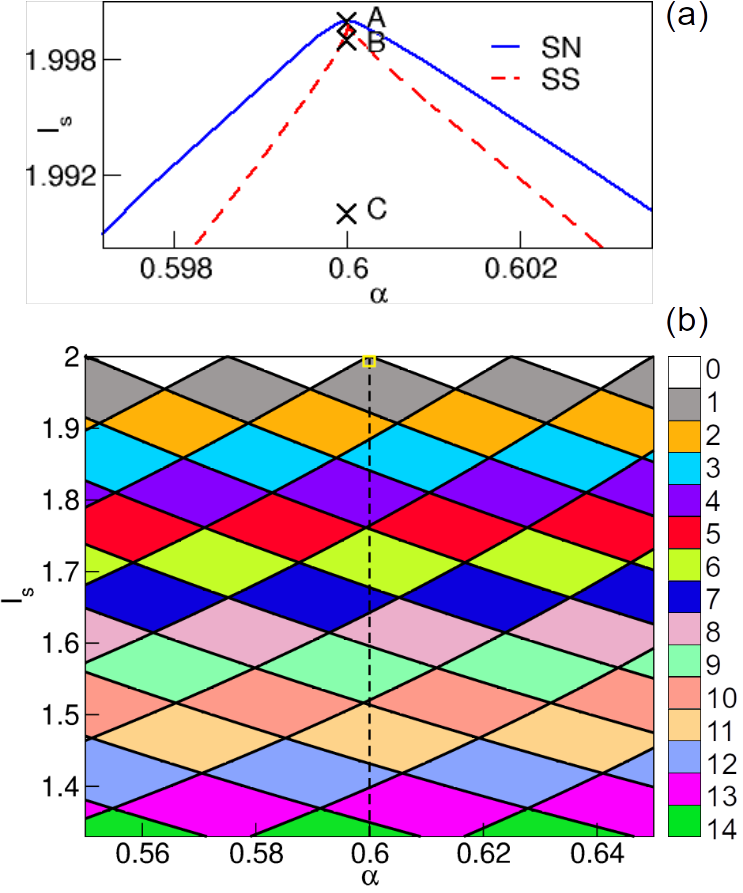}
  \caption{(a) Saddle-node (blue solid line) and saddle-saddle (red dashed line) bifurcation lines occurring in the $(\alpha,I_s)$ parameter plane, in the vicinity of $I_s=2$. Labels A--C mark the $I_s$ values corresponding to Fig.~\ref{fig:nullclines}(a)--(c), respectively. (b) Number of stable fixed points shown with different colors, for $\alpha \in [0.55,0.65]$ and $I_s \in [1.33,2]$. The small yellow box indicates
  the portion of the parameter space depicted in (a). Black solid lines mark the saddle-node bifurcation. Other parameter values: $\beta=4.5$ and $\gamma=10$. Vertical dashed line marks the value 
    $\alpha=0.6$.}
    \label{fig:SN_lines}
\end{figure}

The SN and SS bifurcations which are responsible for the creation of the fixed points of Figs.~\ref{fig:nullclines}(a) and (b)
have been confirmed and followed in the $(\alpha,I_s)$ parameter space. This is done with the help of a very
powerful software tool that executes a root-finding algorithm for the continuation of solutions and bifurcations~\cite{engelborghs2002numerical}. 
The result is shown in Fig.~\ref{fig:SN_lines}(a), where the blue solid and red dashed line denote the SN and SS line, respectively,
and the labels A--C mark the value of $I_s$ corresponding to the fixed points of Figs.~\ref{fig:nullclines}(a--c). 
As the parameters $\alpha$ and $I_s$ are varied further, additional fixed points are born and the bifurcation structure is enriched:
Figure~\ref{fig:SN_lines}(b) shows the number of stable fixed points and the underlying saddle-node bifurcation lines through which they lose their stability.
Note that we have intentionally omitted the SS bifurcation lines because, being in close proximity to the SN lines, they would be hardly distinguishable. 

The borders between different colors mark the annihilation/generation of stable fixed points. For lower values of $I_s$ there are many fixed points, 
while when $I_s>2$ there are none.  
In the following, we will see how the nullclines and the associated fixed points affect the global dynamics and the associated neurocomputation properties
of the system.

\section{Multistable global dynamics}
\label{sec:Dynamics}
\begin{figure}[!htp]
    \centering
    \includegraphics[width=\linewidth]{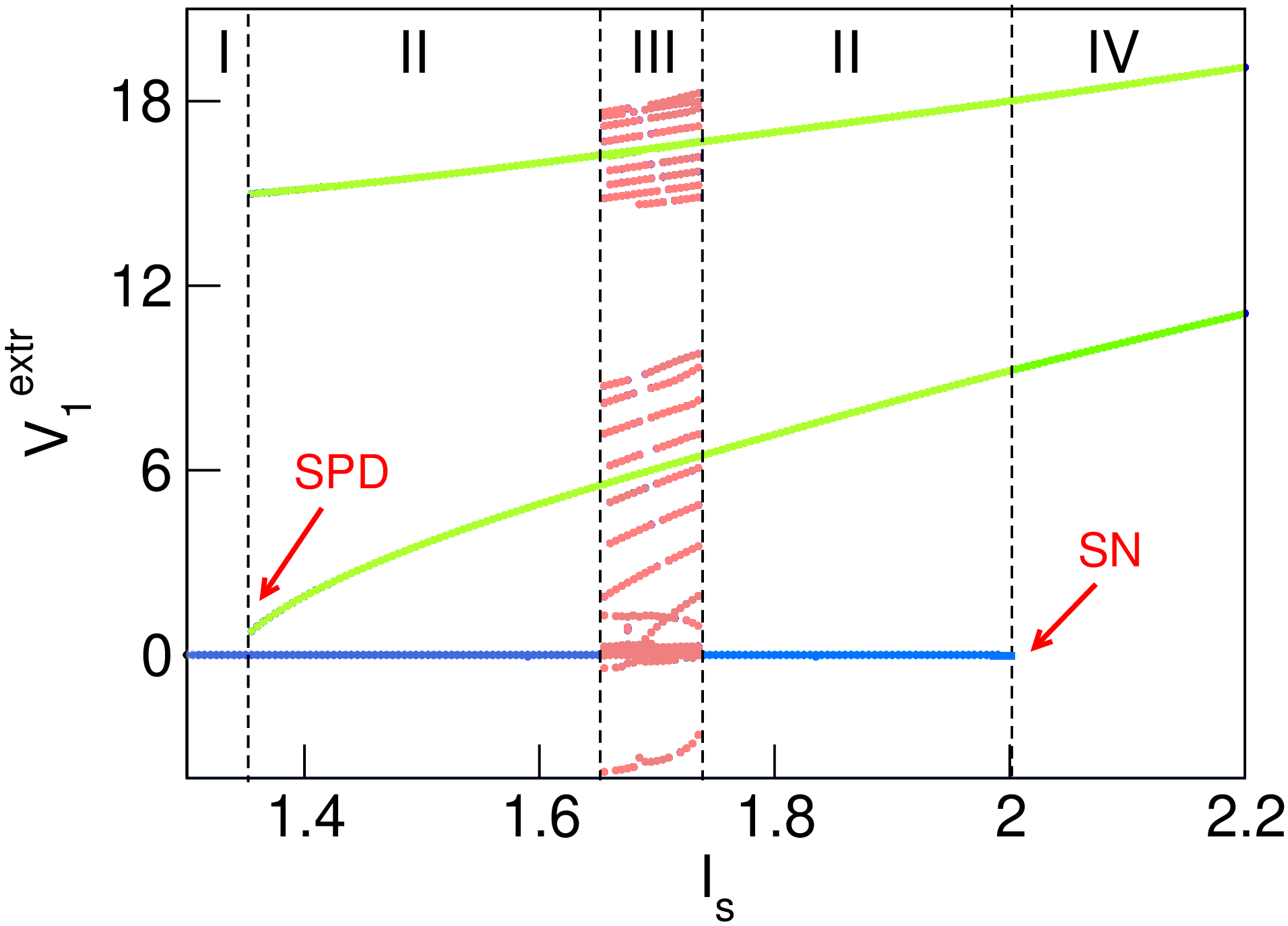}
    \caption{Orbit diagram in terms of the local minima and maxima (extrema) of $V_1$ in dependence of parameter $I_s$. \RNum{1}: the only attractor is a fixed point (neuron resting region), \RNum{2}: coexistence of spiking (light green) and 
     resting (light blue) (bistable region), \RNum{3}: coexistence of bursting (pink), spiking, and resting (multistable region), and \RNum{4}: the only attractor is a limit cycle. ``SPD" marks the subcritical period doubling bifurcation while ``SN" the saddle-node bifurcation.  Other parameters are $\alpha=0.6$, $\beta=4.5$, and 
     $\gamma=10$.}
    \label{fig:orbit_diagram}
\end{figure}

Apart from the fixed points that we analyzed above, the system exhibits additional coexisting solutions
that present very interesting neuron-like properties that we will look into in this section.
For an overview of the full palette of dynamical behavior, we generate the 
orbit diagram in terms of the local minima and maxima of the variable $V_1$ in dependence on the control parameter $I_s$,
shown in Fig.~\ref{fig:orbit_diagram}. In order to capture multistable solutions, for each value of $I_s$ we use different initial conditions 
and, after discarding the transients, the local extrema of $V_1$ are recorded. 

We distinguish four different dynamical regions as $I_s$ is varied: In region \RNum{1} the only attractor is a fixed point and therefore the JJ neuron is resting.
Region \RNum{2} is bistable as there is coexistence of a limit cycle and a fixed point, thus the JJ neuron is capable of both spiking and resting. The limit cycle is born at $I_s=1.3527$ via a subcritical period-doubling (SPD) bifurcation which will be discussed later.
In-between the two bistable regimes there exists a multistable one, region \RNum{3}, where in addition to spiking and resting, the system also 
exhibits bursting dynamics. The multiple points extending over a long range of positive $V_1$ values for a specific $I_s$ correspond to the multiple peaks found within each burst. Finally in region \RNum{4} the only attractor is a limit cycle, since the fixed point disappears via a saddle-node (SN) bifurcation. 

Note that the three coexisting solution branches have been colored according to a palette that is also used in Figs.~\ref{fig:basins} and \ref{fig:Is1.652_3Dplot} for uniformity: Fixed point, limit cycle and bursting solutions
are highlighted with light blue, light green, and pink, respectively.
In what follows, we will undertake an in-depth study of the spiking
and bursting dynamical properties of our system and see how they relate to neurocomputation.

\subsection{Spiking and excitability}
\label{sec:spiking}
Spiking dynamics is observed in the regions \RNum{2} of Fig.~\ref{fig:orbit_diagram} in which there is coexistence with the resting state.
As analyzed in Sec.~\ref{sec:FP} and vizualized in Fig.~\ref{fig:SN_lines}(b), the number of fixed point attractors decreases as $I_s$ approaches the critical value $I_s^\text{crit}=2$.
Naturally, this has an effect on the basin of attraction of the spiking solution which expands as $I_s^\text{crit}$ is approached and shrinks in the opposite direction, in the left region \RNum{2}
of Fig.~\ref{fig:orbit_diagram}. The latter yields it harder to identify the corresponding limit cycle for lower values of $I_s$,
because the phase space contains many equilibria that attract the trajectory of the system, and the spiking solution is less likely to be approached.

\begin{figure}[!htp]
    \centering
    \includegraphics[width=\linewidth]{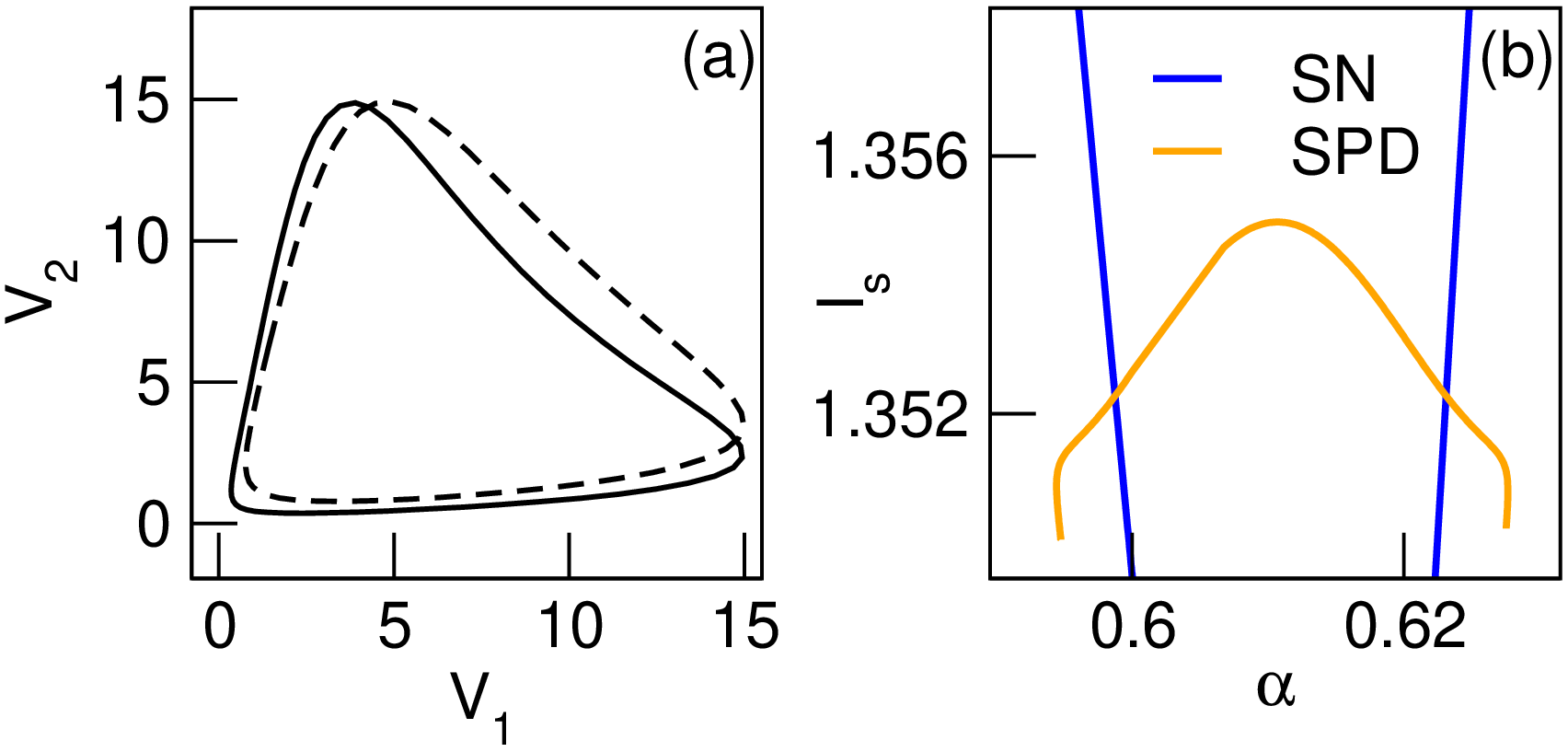}
    \caption{(a) A stable periodic orbit (solid line, $I_s=1.355$) coalesces with an unstable orbit with double its period and loses its stability (dashed line, $I_s=1.345$). Subcritical period doubling (SPD) bifurcation and line and saddle-node (SN) bifurcation line in the $(\alpha,I_s)$ parameter plane. Other parameters are $\alpha=0.6$, $\beta=4.5$, and $\gamma=10$.}
    \label{fig:PD_plot}
\end{figure}

The spiking behavior arises via a subcritical period doubling bifurcation: Viewed from the opposite direction (for decreasing $I_s$)
a stable limit cycle collides with an unstable limit cycle with double its period and loses its stability. Both stable (solid line) and unstable (dashed line) limit cycles are shown in the phase portrait of Fig.~\ref{fig:PD_plot}(a) in the $(V_1,V_2)$ plane for $I_s=1.355$ and $I_s=1.345$, respectively. At the critical point $I_s^{\text{SPD}}=1.3527$ the leading Floquet multiplier crosses the unit circle at the value $-1$ which is a signature of the period-doubling bifurcation and has been confirmed with the bifurcation analysis and continuation software~\cite{engelborghs2002numerical}. The SPD line has also been followed in the $(\alpha,I_s)$ parameter space,
shown in Fig.~\ref{fig:PD_plot}(b) in orange color. In addition we have plotted in blue color the saddle-node bifurcation lines (SN), which have been discussed in detail in Sec.~\ref{sec:FP}.

Strictly speaking, these and all the SN bifurcations that generate the fixed points coexisting with the limit cycle in the Regions \RNum{2} are saddle-node \emph{off} invariant cycle bifurcations.
This is very important because it determines the system's excitability which is classified as type \RNum{2}~\cite{izhikevich2007dynamical}, and is more commonly linked to Hopf bifurcations. This subcategory of type \RNum{2} excitability requires the occurrence of a SN bifurcation and the subsequent ``jumping" of the trajectory to an \emph{already coexisting} limit cycle, which is precisely the case here. As a result, the JJ neuron starts to fire so-called ``tonic" (periodic) spikes of finite period, which have the typical form expected by spiking which is generated via the described bifurcation mechanism, i. e. although they are nonlinear, they appear ``harmonic" [see Figs.~\ref{fig:synch_plot} (a)\& (b) and Fig.~\ref{fig:latency}(a) in the next sections], rather than action-potential-like (with the characteristic de(re)polarization phases and refractory period). Similar spiking behavior has been reported in prior studies on a leech heart interneuron model~\cite{shilnikov2005mechanism}.  

The more typical action-potential-like spikes are associated with type \RNum{1} excitability and a saddle-node bifurcation \emph{on} invariant cycle
(SNIC), alternatively known as saddle-node infinite period (SNIPER) bifurcation~\cite{chalkiadakis2022dynamical}. In fact, our system is indeed capable of producing such spikes, in particular when the parameter $\gamma$ is varied. However, we
have chosen not to address this dynamical behavior in the present work, since the focus here is on the variation of the applied current, which is experimentally feasible to tune, whereas the parameter $\gamma$ is fixed for a given junction and cannot change dynamically.  

It should be noted that different classes of excitability result in different neurocomputational
properties. Specifically, the difference in the spike initiation, as described above,
could potentially affect key brain functions, including information encoding and processing~\cite{izhikevich2007dynamical}.

\subsection{First spike latency}
\label{sec:FSL}
Another interesting behavior that we observe, which is related to the spiking dynamics and its interaction with the fixed points of the system, is the so-called ``first spike latency" (FSL) property.  
The FSL effect is related to the existence of a significant delay in the production of the first spike~\cite{izhikevich2007dynamical}. From a dynamical systems point of view, long latencies 
arise when neurons undergo a saddle-node bifurcation. Since in our system there is a multiplicity of SN bifurcations
and fixed points, we choose to demonstrate the FSL effect slightly above $I_s^\text{crit}=2$, where the last SN bifurcation has just taken place and the only solution is a limit cycle. 

\begin{figure}[!htp]
    \centering
    \includegraphics[width=0.95\linewidth]{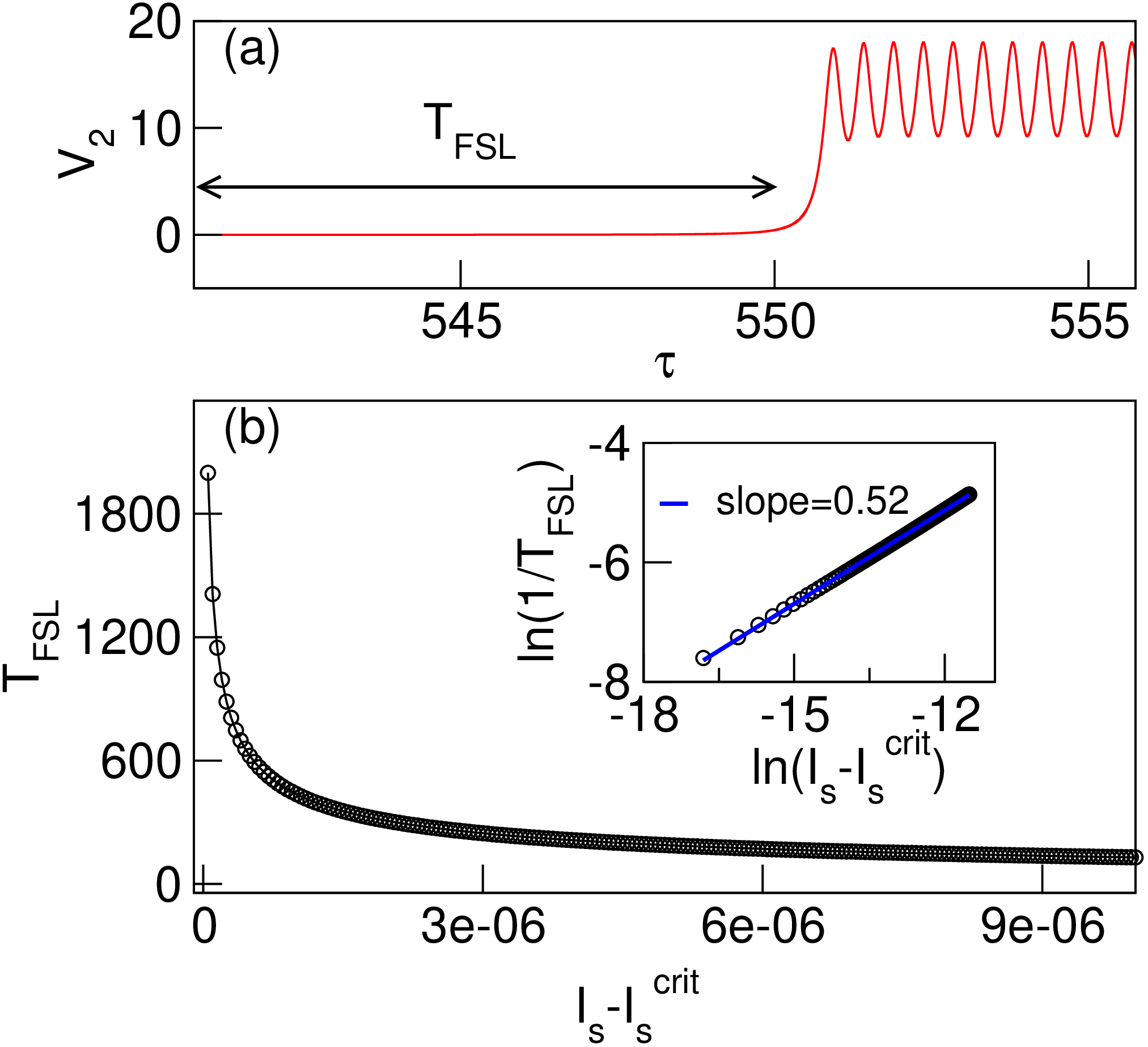}
    \caption{(a) Time-series of the $V_2$ variable showing spike latency. (b) Spike latency time over $I_s-I^\text{crit}_s$, where $I^\text{crit}_s=2$.
    The inset shows the scaling law $1/T_\text{FSL} \sim O(\sqrt{I_\text{s}-I^\text{crit}_s})$ (black circles) and the corresponding linear fit (blue line) with a slope value of $0.52$. Other parameters are $\alpha=0.6$, $\beta=4.5$, and $\gamma=10$.}
    \label{fig:latency}
\end{figure}

Figure~\ref{fig:latency}(a) shows the first spike latency in terms of the voltage variable $V_2$ time series.
We observe a delay
of about $550$ dimensionelss time units before the system ``jumps" to the limit cycle and starts spiking.
This delay is related to the ``bottleneck"  or ``ghost" of the preceding saddle-node bifurcation at 
$I_s^\text{crit}$. More specifically, the system has a ``memory" of the collision of the two fixed points and, when the initial conditions are prepared in its vicinity, the trajectory slows down for a considerable time, before the first spike is fired.   
Note that, due to the fact that the SN bifurcation is \emph{off} an invariant cycle (type \RNum{2} excitability),
once the system starts spiking, it never returns to the bottleneck (and therefore the spiking is of high frequency), 
contrary to the SN on an invariant cycle case (type \RNum{1} excitability), where the trajectory visits the bottleneck after each spike, resulting in ﬁring with a small frequency.

Interestingly, the bifurcation structure is reflected in the bottleneck passage time as a function of the distance from
the critical point. Figure~\ref{fig:latency}(b) shows the first spike latency as a function of $I_s-I_s^\text{crit}$. From the inset, it is evident that the FSL scales as $1/\sqrt{I_s-I_s^\text{crit}}$, which is verified by the near $0.5$ slope of the graph. This behavior has been reported in the bibliography~\cite{izhikevich2007dynamical} and is not to be confused with the square root scaling law of the spiking frequency, observed just above a SNIC bifurcation.

The FSL behavior is closely related to neural encoding, i.~e. the process by which external environmental stimuli are mapped to neuronal electrical activity. In the context of our system, the role of the stimulus is played by the applied current $I_s$. Stimulus information is encoded in spikes, in both the firing rate and the spike timing \cite{kreiman2004neural}. In the first concept, the variable of interest is the amount of action potentials fired, while the second concentrates on the exact time at which spikes take place. 

Concerning spike timing, a significant measure seems to be the time that ensues from stimulus arrival to first spike generation~\cite{gilles2010intrinsic}.
First spike latency seems to be crucial in information integration and relay in biological neurons, especially at the population level in the case of sensory and retinal neurons~\cite{johansson2004first,gollisch2008rapid}. 
Exploiting first spike latency to convey information, leads to a different type of information encoding known as time-to-first-spike coding (TTFS)~\cite{bonilla2022analyzing}. Biological neural systems exploit TTFS, providing them with the significant advantage of a faster and more robust route for information transfer, compared to the firing rate. 
 
In the realm of artificial intelligence, efforts are being made to develop neuromorphic hardware and algorithms that will enable spiking neural networks (SNNs) to perform comparably to classical artificial neural networks (ANNs), with substantially lower energy consumption~\cite{rathi2023exploring}. In~\cite{stanojevic2024high} the authors leverage TTFS coding to achieve equivalent performance to artificial neural networks (ANNs) on standard benchmarks with fewer than $0.3$ spikes per neuron. To do so they overcome the challenges of vanishing or exploding gradients, unstable training dynamics and hardware related fine-tuning constraints, proposing low-latency neuromorphic hardware implementations of deep SNNs with performance on par with ANNs.  

\subsection{Synchronization}
\label{sec:synch}
Up to this point, we have concentrated on the individual behavior of the Josephson junction neurons. By design, however, the system under study is coupled, and as such exhibits interactive behavior too.
Within the spiking region, we will address the issue of phase synchronization between the two Josephson junction neurons. 

\begin{figure}[!htp]
    \centering
    \includegraphics[width=0.95\linewidth]{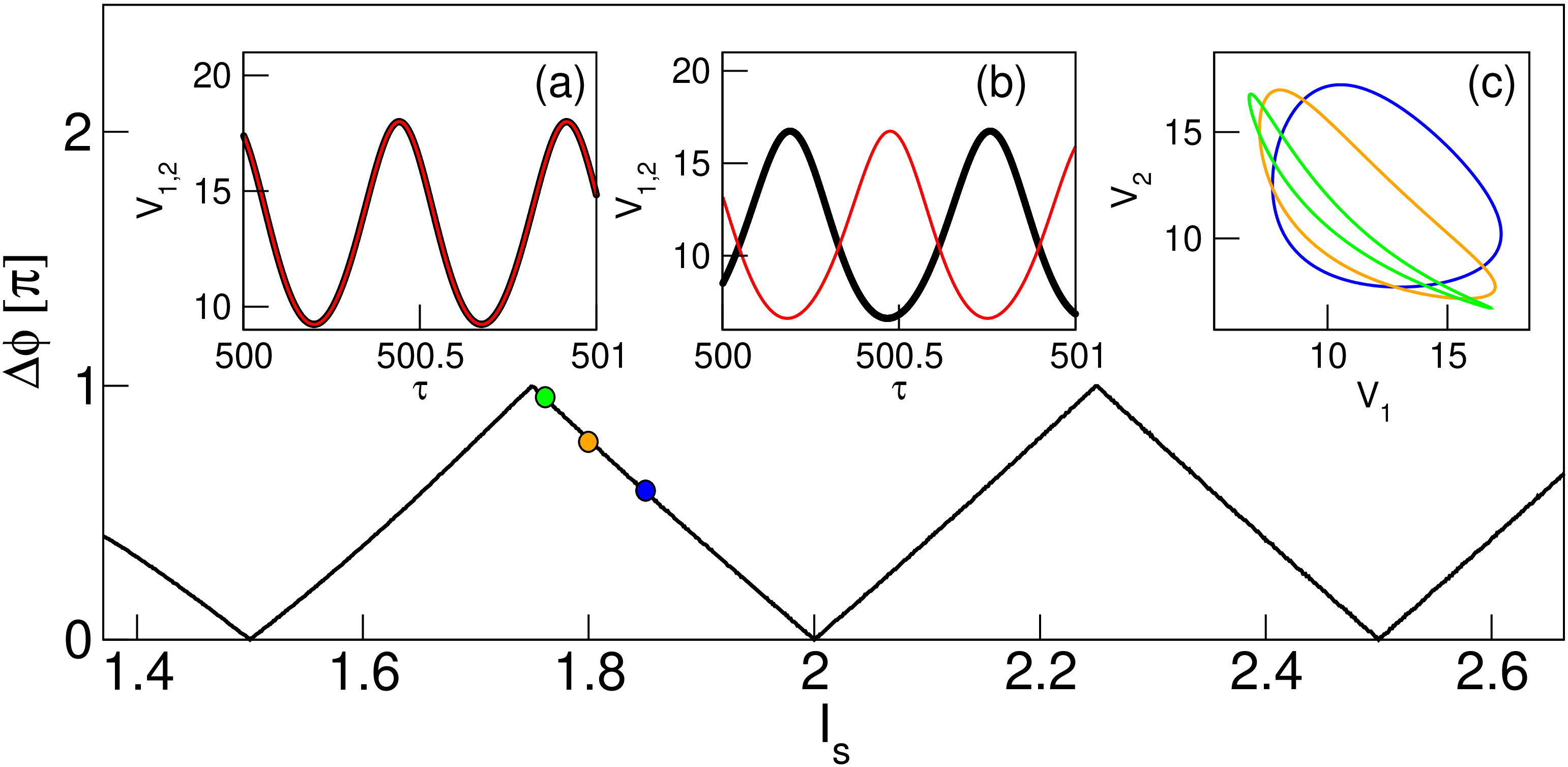}
    \caption{Phase difference $\Delta \phi=|\phi_1-\phi_2|$ of two JJs over $I_s$ for the spiking solution.
    (a) In-phase solution for $I_s=1.75$, (b) anti-phase solution for $I_s=2.0$, (c) Phase portrait for $I_s=1.85$ (blue), $I_s=1.8$ (orange), and $I_s=1.76$ (green).
    The full circles in the $\Delta \phi$ plot have matching colors and mark the values of $I_s$.
    Other parameters are $\alpha=0.6$, $\beta=4.5$, and $\gamma=10$.}
    \label{fig:synch_plot}
\end{figure}

In order to quantify this we calculate the phase difference $\Delta \phi=|\phi_1-\phi_2|$ of the two JJs as a function of the control parameter $I_s$. 
This is plotted in Fig.~\ref{fig:synch_plot} where we observe that the level of synchronization oscillates periodically between the in-phase and
anti-phase state. On the upper limit this is ongoing for $I_s>2.6$ (not included in the figure), while at the lower limit it is terminated
as the limit cycle vanishes in the aforementioned subcritical period-doubling bifurcation. 

The in-phase and anti-phase solutions where $\Delta \phi=1$ and $0$, respectively, are shown in the insets (a) and (b) of Fig.~\ref{fig:synch_plot}, in terms of the voltage variables. Between these two states, we have selected three different values of $\Delta\phi$, marked with colored full circles. The corresponding phase portraits are depicted using the same color code, in the $(V_1,V_2)$ plane of Fig.~\ref{fig:synch_plot}(c), where we see how the trajectory gradually develops from a cycle (lower synchronization, smaller $\Delta \phi$) into a straight line (higher synchronization, larger $\Delta \phi$).

Similar behavior has been observed in a different superconducting circuit
of coupled JJs where a Josephson transmission line acts as the axon and the synapse is modeled by a SQUID. There, it was reported
that the neurons are either desynchronized or synchronized in
an in-phase or antiphase state, and that the tuning of the SQUID 
is capable of switching the system
back and forth in a phase-flip bifurcation~\cite{segall2017synchronization}.

In terms of neurocomputation, synchronization plays a critical role in several cognitive functions by enabling efficient communication and coordination but may also reflect pathological brain states.
In addition, synchronization enhances computational efficiency, eliminating redundant processing and making neural coding more robust and energy-efficient~\cite{fries2005mechanism}.
In our system, synchronization can be easily controlled by tuning the applied current $I_s$. The latter models the input currents arriving at the neuron and may correspond to different synaptic, intrinsic, and external influences. 

\subsection{Bursting}
\label{sec:bursting}
The last region of Fig.~\ref{fig:orbit_diagram} we will examine in this subsection is region \RNum{3},
where spiking and resting coexist with bursting dynamics. In biological neurons, bursting is a rhythmic pattern of spikes, characterized by periods of rapid firing (bursts) interspersed with quiescent intervals. 
Bursting has been found in recordings of real neurons~\cite{izhikevich2007dynamical} and
is considered to be linked to a distinct mode of neuronal signaling~\cite{krahe2004burst}.
Moreover, bursting has profound implications for neurocomputation: 
Its multiscale dynamics, robustness, and efficiency make it essential for tasks involving temporal pattern recognition, learning, and synchronization~\cite{wang2010neurophysiological}.

Dynamically speaking, bursting originates, in general, from a fast motion controlled by a slow motion in a
slow–fast dynamical system. This is in line with what actually happens in biological neurons,
where fast spiking of $Na^+$ and $K^+$ ions are controlled by a slow process
like $Ca^{++}$ gated $K^+$ ion movement.
In autonomous bursting, that is for constant stimulus, there should be an additional variable with a slower timescale than those participating in the spiking, which is responsible for turning off and on the generation of the action potentials~\cite{izhikevich2007dynamical}.

\begin{figure*}[!htp]
    \centering
    \includegraphics[width=\linewidth]{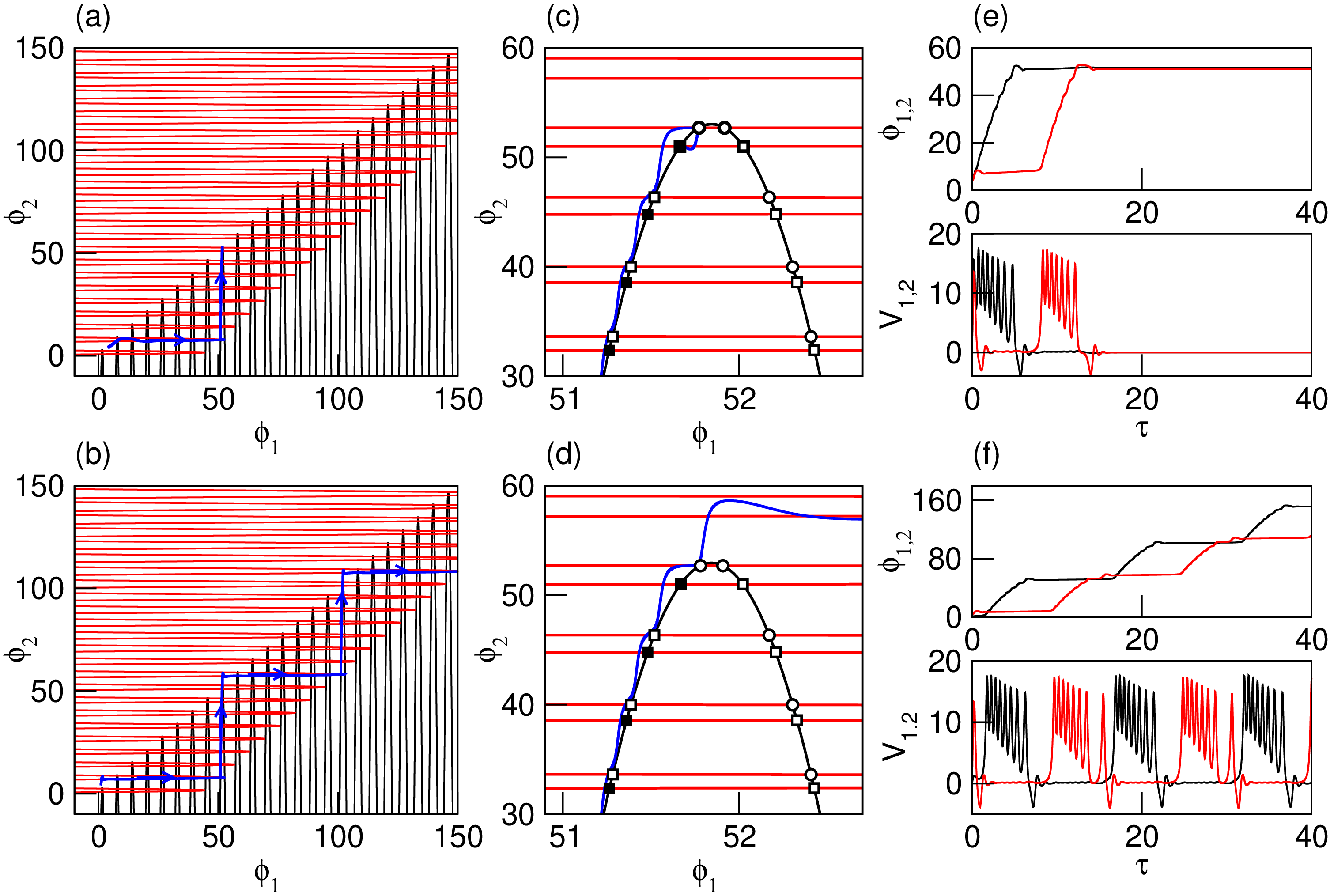}
    \caption{$V_1$ (black) and $V_2$ (red) nullclines (Eqs.~\ref{eq:nullcline V_1} and \ref{eq:nullcline V_2}, respectively) and
    trajectory in the $(\phi_1,\phi_2)$ plane (blue) for (a) $I_s=1.651$ and (b) $I_s=1.652$, and their corresponding close-ups in (c) and (d). Panels (e) and (f) show the system variable time-series for the first (black line) and second (red line) Josephson junction, for $I_s=1.651$ and $I_s=1.652$, respectively. Other parameters are $\alpha=0.6$, $b=4.5$, and $\gamma=10$.}
    \label{fig:Is1.651_Is1.652}
\end{figure*}

Bursting behavior in coupled Josephson junctions was first reported in \cite{blackburn1978dynamics}, however the focus was not on dynamics, let alone neurocomputation, bur rather on the properties of the system as a superconducting interferometer. The mechanism behind both autonomous and non-autonomous bursting has been qualitatively described for the single RCLSJ model in~\cite{dana2006spiking,mishra2021neuron}. The approach is based on the well-known bifurcation scenario of the RCSJ model~\cite{strogatz2018nonlinear}, where spiking is achieved above a critical current threshold via a SNIC bifurcation for higher damping, while for lower damping there exists a bistable regime where resting and spiking coexist.
In this case, spiking is achieved via a homoclinic bifurcation, while the resting dynamics vanishes through a fold (saddle-node) bifurcation at the critical current threshold.
Before we turn to how bursting occurs in our system, we briefly summarize the corresponding bifurcation mechanisms for the single RCLSJ model which follow the classification done in~\cite{izhikevich2000neural}.

Similarly to the RCSJ model, the autonomous RCLSJ system has two fixed points, one saddle point and one stable node, that participate in the aforementioned bifurcation scenario, triggered by the inductive current. For high inductance, the inductive current is slow
enough in comparison to the fast junction voltage. 
When the inductive current starts growing, the junction voltage
starts spiking via a SNIC bifurcation. During
the decaying process of the inductive current, the junction voltage also starts
decaying in a spiral motion into the saddle-point-turned-saddle-focus via a homoclinic
bifurcation. Therefore, the bursting in this case is of SNIC/homoclinic type. 

In the non-automonomous RCLSJ model, there are two mechanisms that generate bursting, depending
on the choice of damping: When the damping parameter is chosen in the SNIC region, 
below the critical current threshold, 
a slow periodic current pushes the dynamics in and out of the spiking regime via a
SNIC bifurcation, generating bursting behavior of SNIC/SNIC type, also known as circle/circle or parabolic type.
Alternatively, if the system is prepared in the bistable region, the external 
driving will periodically force it into the spiking regime 
via a fold bifurcation and bring it back to the resting state 
via a homoclinic bifurcation. In this case the bursting is of fold/homoclinic type.

The fold/homoclinic bursting mechanism has also been attributed to the model  
under study here~\cite{mishra2021neuron}, but merely as an assumption, since no formal study has been conducted to confirm it. In the following analysis, we present a different description
for the creation of bursting in our system.
Besides, this model involves two coupled junctions, i.~e. it is a higher dimensional system
than the single RCLSJ model and, as such, it is expected to exhibit more complex dynamics.

\begin{figure}[h!]
     \centering
     \begin{subfigure}{0.9\linewidth}
         \includegraphics[width=\linewidth]{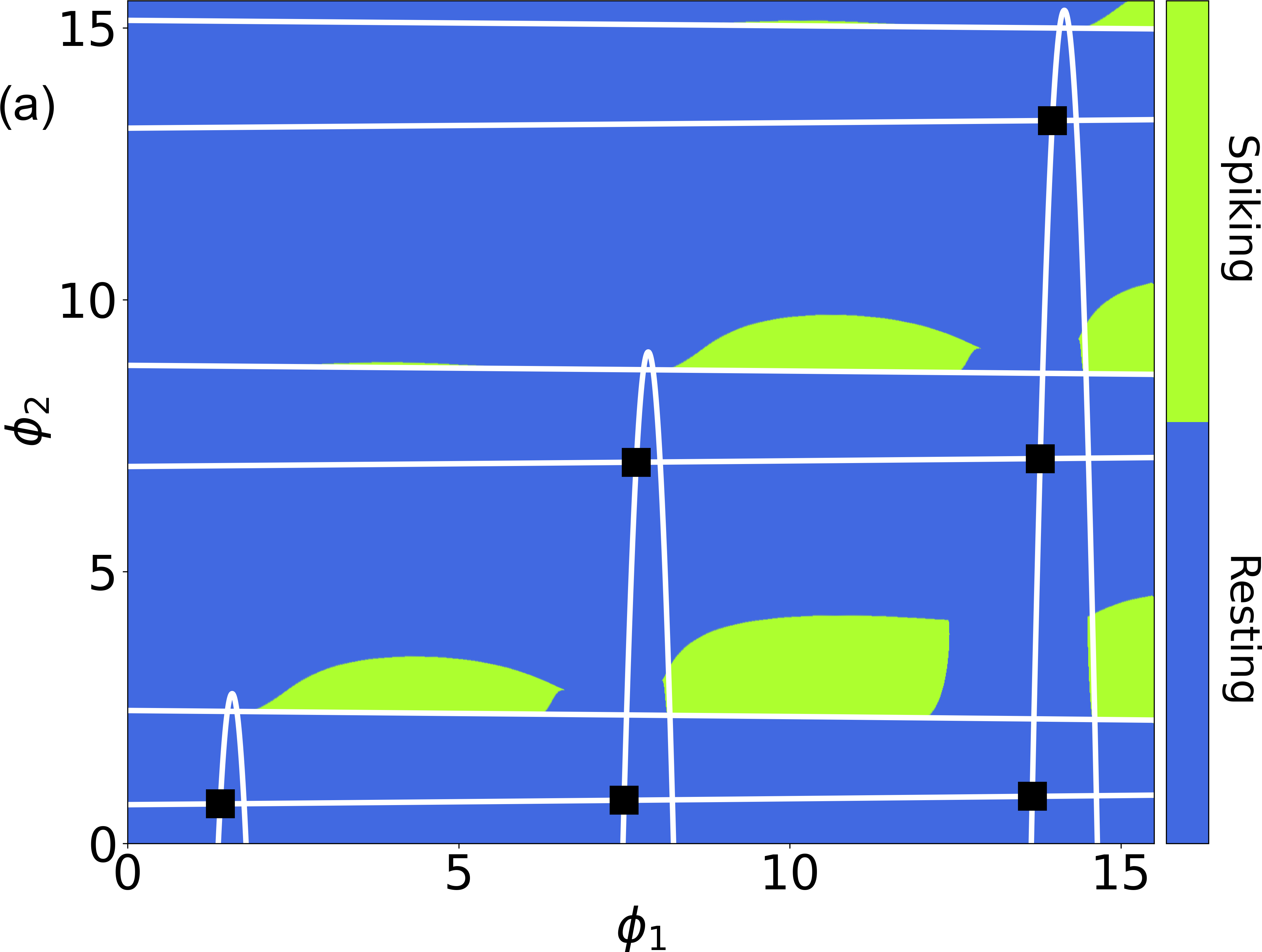}
         \phantomsubcaption{}
        \label{fig:}    
    \end{subfigure}
     \begin{subfigure}{0.9\linewidth}
         \includegraphics[width=\linewidth]{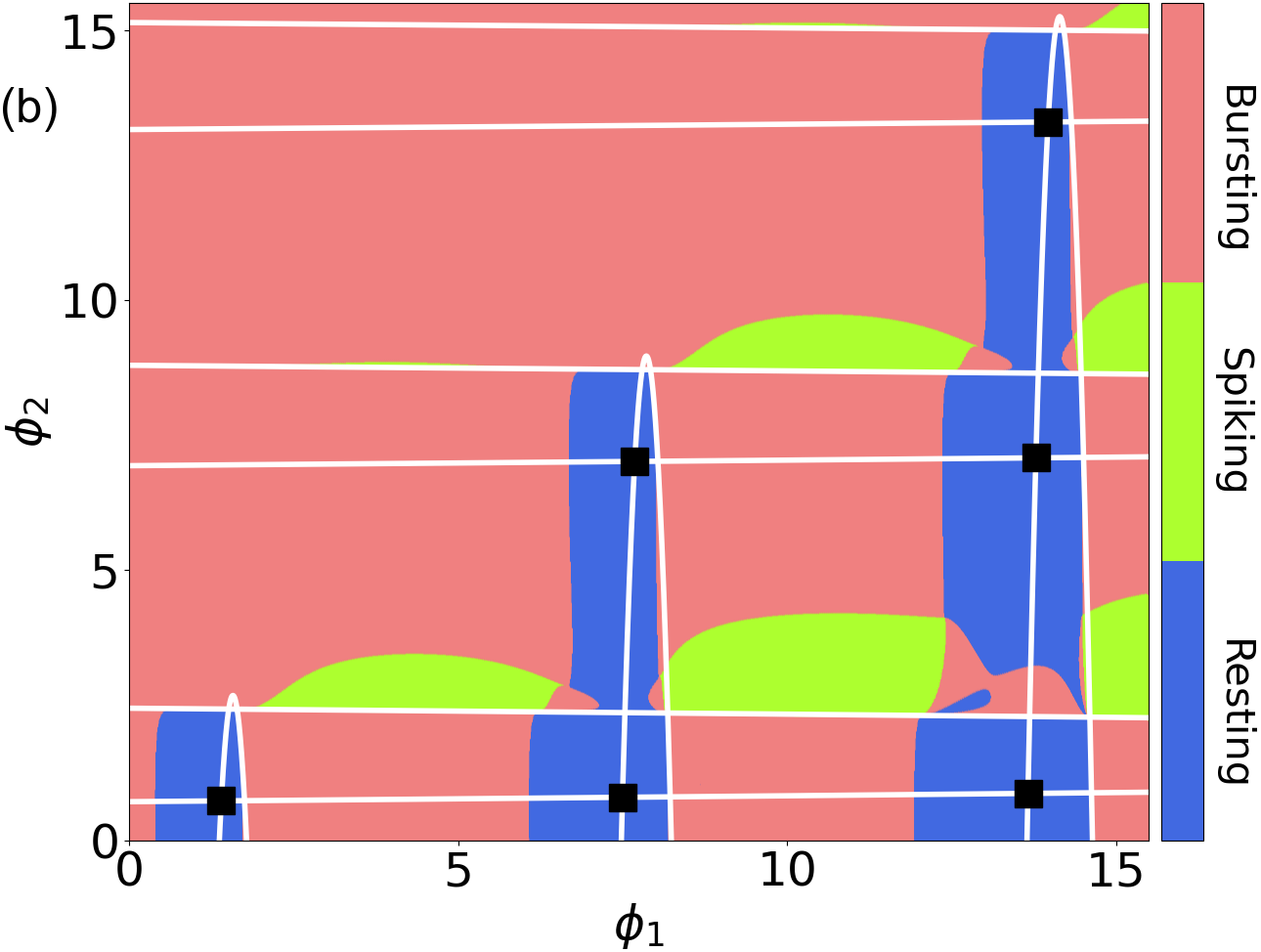}
         \phantomsubcaption{}
         \label{fig:}    
     \end{subfigure}
    \caption{Basins of attraction in the $(\phi_1,\phi_2)$ plane for (a) $I_s=1.651$ and (b) $I_s=1.652$. Initial conditions $V_1=V_2=0$ have been used for the other two variables. 
    Light green, light blue, and pink areas mark spiking, resting, and bursting dynamics, respectively. White lines correspond to the nullclines, while full rectangles to the stable foci. 
    Other parameters are $\alpha=0.6$, $b=4.5$, and $\gamma=10$.}
    \label{fig:basins}
 \end{figure}

As shown in Fig.~\ref{fig:orbit_diagram} the bursting dynamics coexists with spiking and resting behavior (region \RNum{3}). The onset of bursting is observed around $I_s=1.652$ where the system is indeed bistable as required for fold/homoclinic bursting to occur. This bistability is also illustrated in the corresponding basins of attraction in the $(\phi_1,\phi_2)$
plane shown in Fig.~\ref{fig:basins}(a), where light green areas indicate the initial conditions leading to spiking and light blue areas those that lead to resting behavior.
However, contrary to the single RCLSJ, here we have multiple pairs of stable-foci/saddle-foci and saddle/saddle-foci which do not have to undergo a fold bifurcation in order for bursting to occur, as in the case of the RCLSJ model. Neither do we observe a homoclinic transition back into the resting state. 

Once bursting occurs, the basins of attraction change dramatically. 
This is illustrated in Fig.~\ref{fig:basins}(b), where the initial conditions leading 
to bursting are shown in pink. In both of the basins of attraction of Fig.~\ref{fig:basins} we have included the nullclines and the stable foci which are denoted by full rectangles. These are important in order for us to understand how bursting occurs, as explained next.

Just below bursting occurs, for $I_s=1.651$ [Fig.~\ref{fig:Is1.651_Is1.652}(a)], the trajectory is as follows:
The system moves ``along" the $V_2$ nullcline (red nullcline, where $\dot{V_2}=0$) up to its last intersection with the $V_1$ nullcline (black nullcline, where $\dot{V_1}=0$) and then moves ``along" the latter until it is finally attracted by the uppermost stable focus
of this ``slice". This is better vizualized in Fig.~\ref{fig:Is1.651_Is1.652}(c), where a blow-up of the trajectory is shown, as it terminates 
in the stable focus in a spiral motion. The corresponding time-series for all variables are shown in Fig.~\ref{fig:Is1.651_Is1.652}(e). We see
that while the system ``climbs" the $V_2(V_1)$ nullcline, the slow phase variable 
$\phi_1(\phi_2)$ starts growing and at the same time the voltage variable $V_1(V_2)$ starts
spiking. When $\phi_1(\phi_2)$ stops growing, $V_1(V_2)$ starts decaying into the stable focus and the system reaches the resting state.

As soon as bursting bursting takes over, at $I_s=1.652$ [Fig.~\ref{fig:Is1.651_Is1.652}(b)], the first part of the trajectory is similar. The main difference here
is that as the system finishes ``climbing" the $V_1$ nullcline, it is not attracted by the uppermost stable focus, but rather, spends some time in the vicinity of the saddle-point and then manages to escape to the next $V_2$ nullcline [see Fig.~\ref{fig:Is1.651_Is1.652}(d)]
and so on, resulting thus in a constant alternation between spiking and resting, which makes for bursting behavior [Fig.~\ref{fig:Is1.651_Is1.652}(f].
Within the spiking phase, the system is attracted/repelled by the corresponding stable foci/saddle foci along the nullclines. It has no relation to the coexisiting spiking state that is born via a SPD bifurcation, discussed in subsection~\ref{sec:spiking}. This is also a major difference compared to the fold/homoclinic mechanism for bursting observed in the single driven RCLSJ model. Therefore, we believe that the bursting mechanism in the model under study is related to the interaction between the moving nullclines and the multiple fixed points and the way the system navigates through an everchanging phase space.

 \begin{figure}[!htp]
    \centering
    \includegraphics[width=\linewidth]{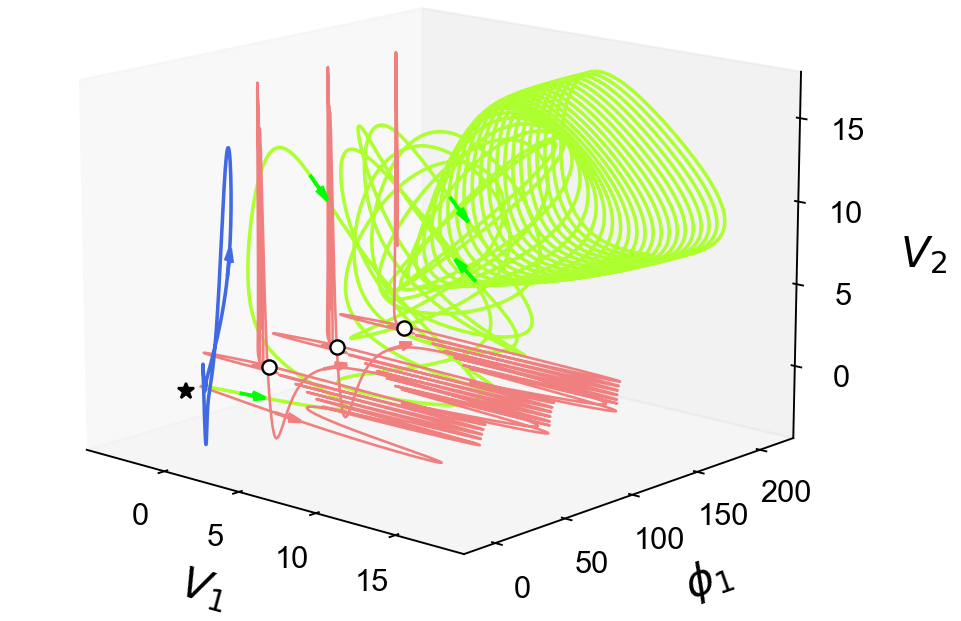}
    \caption{Three-dimensional phase space showing the coexisting trajectories for
    $I_s=1.652$. Light blue, light green, and pink color corresponds to resting, spiking, and bursting, respectively. The star marks the vicinity of the coordinates of the initial conditions used, and open circles denote the saddle-points. 
    Other parameters are $\alpha=0.6$, $b=4.5$, and $\gamma=10$.}
    \label{fig:Is1.652_3Dplot}
\end{figure}

A typical instance of the phase space in the multistable region is finally shown in the three-dimesnional 
projection of Fig.~\ref{fig:Is1.652_3Dplot}. For clarity, we have used the same color code as in Fig.~\ref{fig:Is1.651_Is1.652},
i. e. the resting trajectory is light blue, the spiking one is light green, and the bursting one is in pink color.
All three trajectories start in close proximity (marked by the star). The light blue trajectory is quickly attracted to the stable focus, while the light green one ends up in the limit cycle. The pink trajectory 
oscillates between spiking and resting, the latter of which happens in the vicinity of the saddle points that are also included in the plot.

\section{Conclusions}
\label{sec:Conclusions}
In summary, inductively-coupled Josephson junctions serve as an excellent platform for reproducing key neurophysiological behaviors. This and the fact that they are capable of operating in great speeds with near zero power dissipation yield them very promising candidates for neuromorphic computing.

In this study we addressed a system first introduced in the context of superconducting interferometers
and undertook an in-depth analysis of its neurocomputational properties. 
The model presents multistability where the local dynamics of fixed points (neuron resting state)
interacts with oscillatory motion (neuron spiking and bursting). Via bifurcation analysis
we study the complex ``landscape" involving fixed points and identify the mechanism behind the excitability and the emergence of spiking. In addition, we report on in-phase and anti-phase spiking synchronization and the interchange
between them, as well as first spike latency effects. Particular emphasis has been placed on the study 
of the bursting dynamics exhibited by the system. Unlike previous studies claiming that bursting 
is generated in the same way as in the single resistive–capacitive–inductive shunted Josephson junction model,
in this work we present a different, more elaborate scenario that takes into account the everchanging phase space of the system.

All of the identified dynamics have their biological counterparts and may have implications
in a wide range of neurophysiological behaviors: Bursting is closely related to temporal pattern recognition and learning, synchronization is linked to computational efficiency, and first spike latency plays a crucial role in neural encoding. Further studies could utilize the system for another advancing neuromorphic technology,
namely physical reservoir computing, in which the complex dynamics of physical systems is exploited as
information-processing devices~\cite{tanaka2019recent,Rohm,wang2023moire}. From a merely dynamical point of view, another interesting direction that requires a dedicated mathematical study would be to explore the potential implications of Shilnikov theory~\cite{Shilnikov,shilnikov2005mechanism} involving saddle-foci and homoclinic orbits on our system.

\appendix*
\section{}
\label{sec:appendix}
In the following, we perform a linear stability analysis of the system's fixed points.
The Jacobian of the equilibria given by Eqs.~\ref{eq:nullcline V_1}, \ref{eq:nullcline V_2} reads: 
\begin{equation}
    J = \begin{bmatrix}0 & 1 & 0 &0\\
    -2\pi\gamma \cos\phi_1^*-1/2 &-\beta &1/2&0\\
    0 & 0 & 0 &1 \\
1/2&0&-2\pi\gamma \cos\phi_2^*-1/2&-\beta\\
    \end{bmatrix},
    \label{Jacobian}
\end{equation}
while the characteristic equation is given by:
\begin{eqnarray}
&&\lambda^4+2\beta\lambda^3  +\lambda^2\left[\beta^2+1+2\gamma\pi(\cos\phi_1^*+\cos\phi_2^*)\right]\nonumber \\
&&+\lambda \beta\left[1+2\gamma\pi(\cos\phi_1^*+\cos\phi_2^*)\right ] \nonumber \\
&&+\gamma\pi[\cos\phi_1^*+\cos\phi_2^*+4\gamma\pi\cos\phi_1^*\cos\phi_2^*]=0,
\label{characteristic}
\end{eqnarray}
with roots that provide the eigenvalues of the fixed points:

\begin{align}
    \lambda_1 &= \frac{1}{2}(-\sqrt{-A+C}-\beta) \label{eig_1},\\
    \lambda_2 &= \frac{1}{2}(\sqrt{-A+C}-\beta) \label{eig_2},\\
    \lambda_3 &= -\frac{1}{2}(-\sqrt{A+C}-\beta) \label{eig_3},\\
    \lambda_4 &= \frac{1}{2}(-\sqrt{A+C}-\beta)\label{eig_4},    
\end{align}
where
\begin{align}
    A &= 2\sqrt{4\gamma^2\pi^2(\cos\phi_1^*-\cos\phi_2^*)^2+1}.\\
    C &= \beta^2-4\gamma\pi(\cos\phi_1^*+\cos\phi_2^*)-2.
\end{align}

\end{document}